\documentclass{article}
\usepackage{spconf,amsmath,graphicx,multirow, subfigure}

\usepackage{booktabs}
\usepackage{caption} 
\def \hfillx {\hspace*{-\textwidth} \hfill}

\title{Enhancing ASR for stuttered speech with limited data using Detect and Pass}
\name{Olabanji Shonibare, Xiaosu Tong, Venkatesh Ravichandran}
\address{Amazon Alexa \\
\texttt{olabanjs@amazon.com, tongx@amazon.com, veravic@amazon.com}
}

\begin{document}
%

\maketitle
\begin{abstract}
It is estimated that around 70 million people worldwide are affected by a speech disorder called stuttering \cite{sheikhbahaei2020scientists}. With recent advances in Automatic Speech Recognition (ASR), voice assistants are increasingly useful in our everyday lives. Many technologies in education, retail, telecommunication and healthcare can now be operated through voice. Unfortunately, these benefits are not accessible for People Who Stutter (PWS). We propose a simple but effective method called `Detect and Pass’ to make modern ASR systems accessible for People Who Stutter in a limited data setting. The algorithm uses a context aware classifier trained on a limited amount of data, to detect acoustic frames that contain stutter. To improve robustness on stuttered speech, this extra information is passed on to the ASR model to be utilized during inference. Our experiments show a reduction of 12.18\% to 71.24\% in Word Error Rate (WER) across various state of the art ASR systems. Upon varying the threshold of the associated posterior probability of stutter for each stacked frame used in determining low frame rate (LFR) acoustic features, we were able to determine an optimal setting that reduced the WER by 23.93\% to 71.67\% across different ASR systems.
\end{abstract}
\begin{keywords}
Stuttered speech, Disfluencies, Limited data, Recurrent neural network transducer, End-to-end speech recognition
\end{keywords}
\section{Introduction}
\label{sec:intro}

Automatic Speech Recognition (ASR) systems have been trained on millions of hours of data \cite{parthasarathi2019lessons} and have proven to be robust to various forms of speech, accents, dialects and presence of background noise. Recent trends in speech processing technology powered by neural network frameworks like Recurrent Neural Network Transducer (RNN-T) \cite{alexrnnt} indicate the ability for the model to generalize across various forms of speech. Like most Artificial Intelligence (AI) systems, this technology is subject to majoritarian bias and under-performs for speakers with disfluent speech patterns \cite{morris2020ai}. 

There is a wide spectrum of speech disfluencies that are observed out of which stuttering appears to be the most common one \cite{Kourkounakis}.  Stuttering, which is also referred to as stammering or dysphemia is a disorder that impedes the smooth production of speech sounds. Typical symptoms of stuttering includes long silent pauses between sounds and words (also known as a block), repetition or prolonged pronunciation of certain sounds, syllables or words. The forms of disfluencies \cite{shriberg1994preliminaries} considered in this work include Revision, Interjection, Dysrhythmic phonation, Block, Phoneme repetition, Part-word repetition, Word repetition and Phrase repetition.

Speech recognition systems like Recurrent Neural Networks Transducer (RNN-T) or models based on Connectionist Temporal Classification (CTC) \cite{graves2006connectionist}, that have been trained on many hours of fluent speech, tend to have majoritarian bias as access to speech samples from PWS is limited. The motivation behind this work is to evaluate the relative Word Error Rate Reduction (WERR) for stutterred speech as compared to an ASR model, by using a simple algorithm that uses a classification model trained on limited data to detect frames that contain stutter and pass them to the state of the art ASR inference engine.

In the next section, we describe a summary of the prior work done in the area. We explain the Detect and Pass method in Section 3. Section 4 summarizes the data collection process. Experiments and associated results are discussed in Section 5, which is followed by our conclusion in Section 6.


\section{RELATED WORK}
\label{sec:prior}
\begin{figure*}[h]
    \centering
    \includegraphics[scale=0.63]{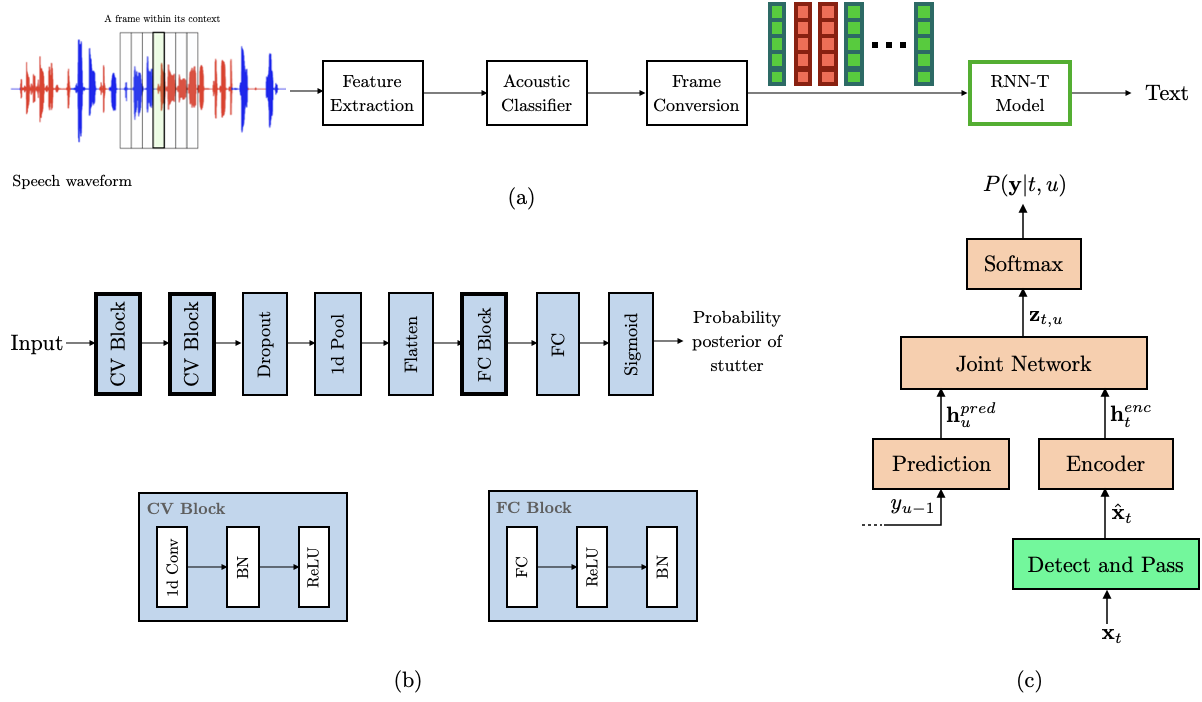}
    \caption{(a) Overview of Detect and Pass Module (b) Acoustic classifier (c) Modified RNN-T model with Detect and Pass module.}
    \label{fig:methods2}
\end{figure*}
Speech disfluencies is a term that is used to discuss a spectrum of speech related irregularities observed in audio data collected from diversity of speakers. Across the published literature, \cite{mcdougall2017profiling} we have observed organic disfluencies (like part word repetitions or hesitations) during conversational speech from speakers who speak fluently. Stutterred speech and dysarthric speech as experienced by people who suffer from conditions like Aphasia, represent the other extreme of the speech disfluency spectrum. Valentin et. al. \cite{Valentin} studied the robustness of RNN-T based ASR models on disfluent speech that contained organic disfluencies like partial words using filters on utterance transcriptions that are indicative of hesitations and repetitions. We introduce the term organic disfluency to distinguish the speech containing hesitations and repetitions from people who do not self identify as People Who Stutter or have not been diagnosed with speech disorders. The study demonstrated a 16.1\% WER reduction (WERR) relative to baseline RNN-T model (trained on fluent speech), on stuttered speech. Colin et. al. \cite{Colin_apple} used a ConvLSTM approach that learns how acoustic features should be weighted using a weighted cross entropy term with focal loss and concordance correlation coefficient (CCC) to show a 16\% WERR on the Fluency Bank dataset.

Stutter is observed in 5\% to 10\% of children's speech , who are aged between 2 and 6 years \cite{NIDCD}. Sadeen et. al. \cite{Sadeen_children} proposes a lightly supervised approach to detect stuttering in children's speech, reducing the relative WER by 27.8\% as compared to baseline ASR model which uses a statistically trained Language Model based only on the original prompt (tends to delete the stuttering events in the transcription). They demonstrated that ASR systems with task-specific re-scoring greatly increased the detection of stuttering events except prolongation. For stutter events related to prolongation, they introduced a correction layer that utilized the correlation between successive frames.

Speech corrections systems have been explored in the past where disfluencies in speech have been eliminated to provide a fluent speech output to downstream language processing \cite{liu2005comparing} \cite{maskey2006phrase} \cite{honal2003correction} \cite{snover2004lexically}. A key component in speech correction systems are classifiers that are able to detect and classify speech as stuttered speech. Shakeel et. al \cite{Shakeel} introduced a StutterNet model that detects stutter using a time delay neural network (TDNN) on UCLASS dataset \cite{howell2009university}, improving the overall average detection accuracy by 4.69\% as compared to ResNET+BiLSTM \cite{kourkounakis2020detecting} baseline model. Sadeen et. al \cite{Sadeen_seq} studied two approaches to do sequence labeling for stuttered speech, Conditional Random Fields (CRF) and Bi-directional Long Short Term Memory Networks (BI-LSTM). The study showed an improvement in F1 Score by 33.6\% for the BiLSTM model relative to CRF baseline, using ngram based features. A further improvement of 45\% was observed when CRF based classifier utilized word and character based distance metrics as input features.

\section{Detect and Pass}
\label{sec:methods}

Our approach to re-mediate a stuttered utterance starts by splitting the utterance into non-overlapping frames. The log mel-spectrogram of each frame is then passed to a trained classifier to determine if it contains stutter. Each frame together with a metadata indicating whether it contains stutter is sent to the ASR model. The ASR model then utilizes the extra information during its decoding process and skips frames with stutter. This process is summarized in Figure~\ref{fig:methods2}(a).

Each audio recording, sampled at 16KHz, is split into non-overlapping frames of length 100 ms. The features that correspond to a frame and its label constitute an instance in the training set. In this work, we take into consideration the context of a frame when computing its features. We found four neighboring frames to the left and right of the current frame to be optimal. These nine frames are then concatenated in order from left to right.

The merged audio frames are fed to the feature extraction module which then computes the log mel-spectrogram. This was obtained using a Short-time Fourier Transform parameterized  by a frame size = 25ms, hop size = 12.5 ms, 512-point FFT and a Hamming window, for each merged frame. The features that correspond to a frame and its label constitute an instance in the training set. The computed features is passed to the neural network model to determine the posterior probability. A stutter event is said to be present in the frame if the posterior is at least 0.5 while a value less than 0.5 signifies absence of stutter. Our neural network model consists of 2 Convolutional (CNN) blocks and 2 Fully-connected (FC) layer. Each CNN block contains a 1D convolutional layer, a batch normalization layer and ReLU activation layer in that order. The first FC layer consist of a linear layer, followed by ReLU activation and Batch normalization layer. The second consists of a linear layer, followed by a sigmoid activation layer.

The RNN-T model, shown in Figure~\ref{fig:methods2}(c), constitutes three networks - Encoder ($\mathcal{F}^{enc}$), Prediction ($\mathcal{F}^{pred}$) and a Joint network, ($\mathcal{F}^{joint}$). The encoder network plays the role of an acoustic model, the prediction network performs a similar role as a language model while the joint network carries out the task of aligning the input and output sequences. In Figure~\ref{fig:methods2}(c), $\mathbf{x}_{t}$ represents the acoustic feature vector and it includes an extra field that contains a prediction from the classifier (whether or not it contains stutter). For each time $t$, $\mathbf{x}_{t}$ is passed onto $\mathcal{F}^{enc}$ to produce a high-level representation $h_{t}^{enc}=\mathcal{F}^{enc}(x_{t})$. $\mathcal{F}^{pred}$ takes as input its previous non-blank output, $y_{u-1}$ to produce $h_{u}^{pred} = \mathcal{F}^{pred}(y_{u-1})$, where $u$ is the index used for the output labels. The joint network is a fully-connected layer that takes as input the output of the encoder network $h_{t}^{enc}$ and the prediction network $h_{t}^{pred}$ to estimate the label distribution at the location u of the output as follows

\begin{align}
    z_{t,u} &= \mathcal{F}^{joint}(h_{t}^{enc},h_{u}^{pred}) \\
            &= \phi(W_{zt}h_{t}^{enc}+W_{zu}h_{u}^{pred}+b_{z})
\end{align}
where $W_{zt}$ and $W_{zu}$ are weight matrices, $b_{z}$ is a bias vector, and $\phi$ is a non-linear activation function like ReLU or tanh. The final posterior for each output token $k$ is then obtained as

\begin{align}
    h_{t,u} &= W_{h}z_{t,u}+b_{h} \\
    P(k|t,u) &= softmax(h_{t,u}^{k}).
\end{align}
where $W_{h}$ is a weight matrix and $b_{h}$ is a bias vector.

\section{Datasets}
\label{sec:datacollect}
Speech samples were collected from people who self-identified as People Who Stutter, through external vendors. The data was de-identified by the external vendors prior to the experiments to preserve privacy. As we were interested in speakers that show a significant amount of stutter, each speaker was asked to read a set of pan-grams that were carefully selected to induce stutter. The severity of stutter for each speaker was then estimated from these audio recordings. Similar to this work [5], we defined the severity of stutter to be the proportion of syllables repeated or prolonged while talking. If stuttering is seen for less than 1\% of syllables, it is considered Normal; 1-6\% is classified as Mild; 6-13\% is classified as Moderate; 13-20\% is classified as Severe and if more than 20\% of syllables is stuttered on, we group as Very Severe. All speakers shortlisted for this study show at least Moderate stuttering. In addition, 3 speakers showed mild; 31 speakers, moderate; 27 speakers, severe and 33 speakers demonstrated very severe form of stuttering.

\begin{figure}[h]
    \centering
    \includegraphics[width=0.48\textwidth]{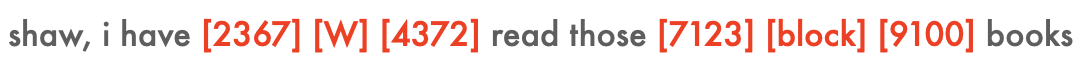}
    \caption{Annotation of an utterance with stutter.}
    \label{fig:annotation2}
\end{figure}

For each audio recording, the onset and offset of a stutter event, in addition to the kind of stutter were manually labeled. An example annotation is shown in Figure~\ref{fig:annotation2}.
The signature [2367] [W] [4372] indicates the stutter type with tag W (word repetition for this work) was observed from times 2367 ms to 4372 ms. On the whole, we had gathered 21 hours of data from 94 unique speakers, comprising a total of 17K utterances.

\section{Experiments}
\label{sec:Experiments}

\subsection{Experimental setup}
As our baseline, we use an RNN-T model. Within the RNN-T setup, we use different variations of RNN-T models trained on different sizes of data sets, varying in the number of hours of training data. The classifier model is trained on a limited training data set of stuttered speech. Our training data set is 16.8 hours as compared to $>10000$ hours of data that baseline ASR models are trained on. Our intent in this work, is not to demonstrate a highly accurate classifier. Instead we want to build a good-enough classifier and demonstrate that this can achieve significant WER reduction for low resource customer cohorts like Stuttered Speech.
For purposes of our experiments, we used a test data set that contained 1725 speech samples (2.5 hours of audio) from People Who Stutter. The data set contained 44 samples of mild stutter, 364 samples of moderate stutter, 336 samples of severe stutter and 981 samples of very severe stutter.

\begin{table*}[t]
\begin{minipage}{0.5\textwidth}
  \caption{WER Reduction (WERR) for Detect and Pass relative to baseline, on state of the art ASR models. $\text{WERR} = 100*(m-b)/b$, where $m$ is the corresponding model WER and $b$ is WER of the baseline model on the same test data set.}
  \label{table:2}
  \setlength{\tabcolsep}{0.5\tabcolsep}
  \centering
  \begin{tabular}{r | c c }
    \toprule
    \textbf{ASR Model} & \textbf{Training data size} & \textbf{WERR (\%)} \\
    \midrule
    RNN-T1  & $200$K hours & 71.24 \\
    RNN-T2  & $500$K hours & 38.03 \\
    RNN-T3  & $800$K hours & 12.18 \\
    \bottomrule
  \end{tabular}
\end{minipage}
\hfillx
\begin{minipage}{0.5\textwidth}

  \caption{WERR (\%) relative to baseline for different RNN-T models.}
  \label{table:3}
  \setlength{\tabcolsep}{0.5\tabcolsep}
  \centering
  \begin{tabular}{r | c c c}
    \toprule
    \textbf{Approach} & \textbf{RNN-T1} & \textbf{RNN-T2} & \textbf{RNN-T3} \\
    \midrule
    majority  & 71.22 & 38.03 & 12.18\\
    any\_1 & 69.58 & 34.34 & 7.66\\
    any\_0  & \textbf{71.67} & 40.05 & 17.26\\
    ave\_0.2  & 43.98 & -20.04 & -77.22\\
    ave\_0.4  & 66.59 & 27.16 & -3.01\\
    ave\_0.5  & 70.89 & 38.55 & 13.61\\
    ave\_0.6  & 69.73 & \textbf{44.85} & 22.56\\
    ave\_0.8  & 70.83 & 43.39 & \textbf{23.93}\\
    ave\_0.9  & 54.15 & 28.29 & 19.05\\
    \bottomrule
  \end{tabular}
\end{minipage}
\end{table*}

\subsection{WER improvements}
Annotation of a stuttered speech requires the annotator to patiently listen to speech and mark the incidence of stutter, with its respective start and end time. This process is prone to annotation errors and labels can be very noisy. For the purposes of this study, we ignored the types of stutter and focused on building a binary classifier. Given a limited data setting, we define a good enough classifier to have Precision-Recall Area Under the Curve (PR-AUC) \cite{flach2015precision} to be greater than 75\% for a test data set. 

\begin{figure}[h]
    \begin{flushleft}
    \includegraphics[width=0.48\textwidth]{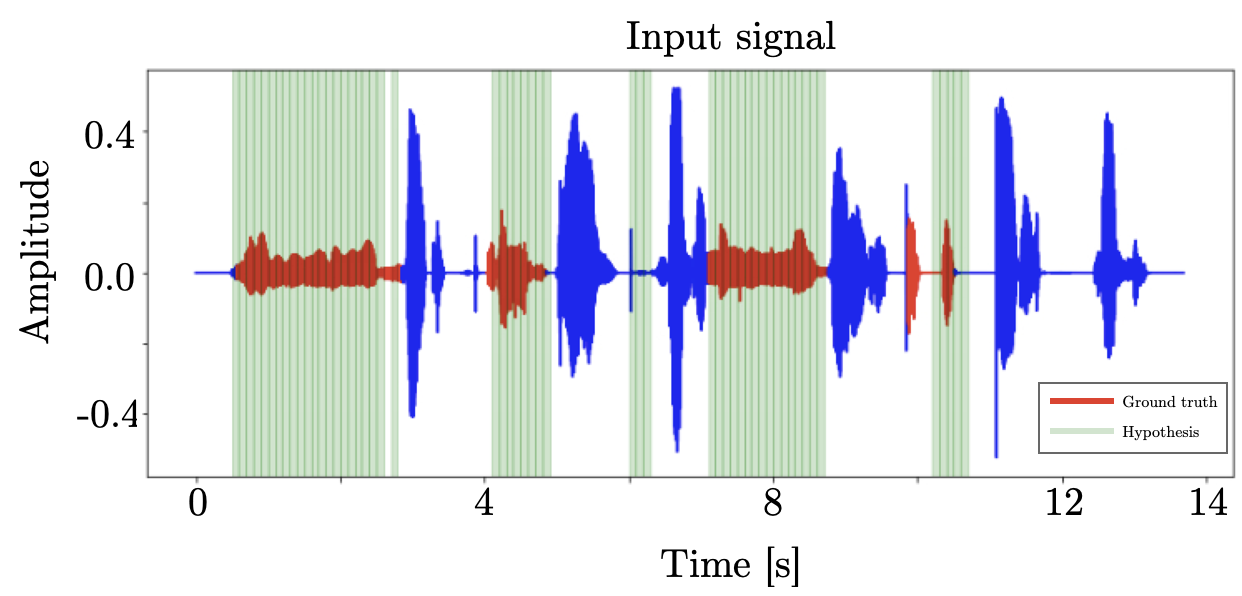}
    \caption{ Comparison of our model prediction (green) and ground truth (red) for a random audio file in the test set }
    \label{fig:methods1}
    \end{flushleft}
\end{figure}

A qualitative impression of our classifier for a randomly chosen speech sample from the test set is shown in Figure~\ref{fig:methods1}. The portions of the speech waveform in blue and red represent fluent and disfluent (stutter) regions of the speech respectively. As discussed earlier, the waveform is split into non-overlapping 100 ms frame. Each frame is then sent to the classifier for prediction. The green rectangles in the figure indicate stutter regions by the classifier and each has a length of 100ms. We observe the model is able to detect almost all regions of stutter for this audio.



In the limited data setting, we evaluated the Detect and Pass technique with baseline ASR models. The results are shown in Table \ref{table:2}. RNN-T1, RNN-T2, RNN-T2 represent different variants of the RNN-T model that (differ) in the amount of training hours. We observed that the Detect and Pass approach is more beneficial to models that are trained on relatively smaller data sets. As ASR models generalize over larger data sets, the relative WER reduction decreases. One positive surprise from experiments is the impact a classifier (trained on a relatively small data set) has on ASR models. This method demonstrates an additive approach to make existing ASR models robust to a specific cohort by using the proposed Detect and Pass technique, in a limited data setting.


\subsection{LFR experimentation}
As Low Frame Rate (LFR) models are popular mechanism to reduce latency in an ASR model [18], we experimented with different frame conversion methods for the Detect and Pass approach. For this evaluation, we define an LFR frame as a stack of three consecutive acoustic frames. We then evaluate different thresholds as well as other voting  techniques  to  determine an optimal WERR setting.


\begin{table}[h]

\end{table}

An assessment of the impact of different frame conversion methods is delineated in Table~\ref{table:3} for RNN-T1, RNN-T2 and RNN-T3. For each approach, we have the following situations: majority - majority vote; any\_1 - If stack contains any stutter frame (1), assign 1, otherwise 0; any\_0 - If stack contains any regular frame (0), assign 0, otherwise 1; ave\_th - if average posteriors $>$ th, assign 1, otherwise 0. We observe any\_0 performs better for RNN-T1 and as the size of the model increases from RNN-T2 to RNN-T3, ave\_0.6 and ave\_0.8 respectively show least impact. Perhaps, as the size of the model increases beyond RNN-T3, ave\_0.8 remains optimal and no further improvement is seen.

\section{Conclusion}
\label{sec:Conclusion}
Through this work, we demonstrate that a relatively simple technique of detecting the stutter frame with a good-enough classifier trained on limited data and passing those frames (and the corresponding posteriors) to the ASR model, reduces the overall WER by a significant margin. This margin is larger when ASR models are trained on relatively smaller data sets. As ASR models are trained on newer technology and larger data sets, this relative WERR becomes smaller. Hence we conclude that the Detect and Pass technique is a valuable and simple method to reduce WER, in a limited data setting.

\vfill\pagebreak

\bibliographystyle{IEEEbib}
\bibliography{strings,refs}

\end{document}